\documentclass[12pt]{iopart}
\usepackage[T1]{fontenc}
\usepackage[latin9]{inputenc}
\usepackage{color}
\usepackage{verbatim}
\usepackage{textcomp}
\usepackage{graphicx}
\usepackage{setspace}
\usepackage{subscript}
\usepackage[english]{babel}

\begin{document}

\title{Compared growth mechanisms of Zn-polar ZnO nanowires on O-polar ZnO
and on sapphire}

\author{G. Perillat-Merceroz$^{1,2}$\footnote{Corresponding author: gperillat@gmail.com}
, R. Thierry$^1$, P.-H. Jouneau$^2$, P. Ferret$^1$ and G. Feuillet$^1$}

\address{$^1$ CEA, LETI, Minatec Campus, Grenoble, 38054, France\\$^2$ CEA INAC/UJF-Grenoble1 UMR-E, SP2M, LEMMA, Minatec Campus, Grenoble, 38054, France}

\begin{abstract}
Controlling the growth of zinc oxide nanowires is necessary to optimize
the performances of nanowire-based devices such as photovoltaic solar
cells, nano-generators, or light-emitting diodes. In this view, we
investigate the nucleation and growth mechanisms of ZnO nanowires
grown by metalorganic vapor phase epitaxy either on O-polar ZnO or
on sapphire substrates. Whatever the substrate, ZnO nanowires are
Zn-polar, as demonstrated by convergent beam electron diffraction.
For growth on O-polar ZnO substrate, the nanowires are found to sit
on O-polar pyramids. As growth proceeds, the inversion domain boundary
moves up in order to remain at the top of the O-polar pyramids. For
growth on sapphire substrates, the nanowires may also originate from
the sapphire / ZnO interface. The presence of atomic steps and the
non-polar character of sapphire could be the cause of the Zn-polar
crystal nucleation on sapphire, whereas it is proposed that the segregation
of aluminum impurities could account for the nucleation of inverted
domains for growth on O-polar ZnO.

\end{abstract}

\section{Introduction}

ZnO nanowires are easily grown, with no catalysts, by different growth
methods on various types of substrates such as sapphire \cite{park_metalorganic_2002},
ZnO \cite{thiandoume_morphology_2009}, GaN \cite{hong_position-controlled_2007},
silicon \cite{lee_comparative_2005}, metal \cite{park_catalyst-free_2008},
graphene \cite{kim_vertically_2009}, and glass \cite{lee_comparative_2005}.
Consequently, they are considered for many applications. For example,
nanostructures with a high surface to volume ratio can be deposited
onto low cost substrates, through cheap growth methods such as chemical
bath deposition: when covered with a thin photon absorber, these nanostructures
open the way to photovoltaic applications \cite{ko_nanoforest_2011,consonni_synthesis_2011}.
Type II hetero-structures based on ZnO nanowires are also envisaged
to absorb the photons and separate the electron-hole pairs \cite{wu_all-inorganic_2011}.
Another possible application field is the fabrication of nano-generators
using the piezoelectric effect of ZnO nanowires in order to convert
mechanical energy into electrical energy \cite{xu_self-powered_2010}.
ZnO nanowires are also studied for applications in light-emitting
diodes (LEDs) or lasers, but it is difficult to dope ZnO \emph{p}-type.
However, recent publications are encouraging concerning the feasibility
of \emph{p}-type doping \cite{nakahara_nitrogen_2010,lautenschlaeger_model_2011,kato_impact_2011}.
Concerning nanowire LEDs, a weak ultra-violet emission was observed
under current injection for \emph{p}-\emph{n} homo-junctions in ZnO
nanowires \cite{willander_zinc_2009-1,chen_near_2010}. A more intense
electroluminescence was obtained from hetero-junctions made of \emph{n}-type
ZnO nanowires coated with another \emph{p}-type material, for example
\emph{p}-type GaN \cite{lupan_low-voltage_2010} or polymers \cite{willander_zinc_2009}.
Finally, an electrically-pumped laser made of Sb-doped \emph{p}-type
ZnO nanowires on a \emph{n}-type ZnO thin film was demonstrated \cite{chu_electrically_2011}.
Using ZnO nanowires instead of two-dimensional ZnO layers presents
definite advantages for engineering LEDs. Indeed, nanowire growth
on large and conductive hetero-substrates such as silicon and metallic
substrates is possible, and no structural defects are expected (as
observed for nanowire growth on sapphire, where no structural defects
were observed \cite{rosina_morphology_2009,perillat-merceroz_mocvd_2010}).
Furthermore, core-shell quantum wells with a large developed surface,
can be grown \cite{bae_fabrication_2006,cao_homogeneous_2009,thierry_core-shell_2012}.
Finally, light extraction is naturally more efficient in a nanowire-based
LED \cite{henneghien_optical_2011}. 

In order to reach these application goals, a controlled growth of
the nanowires is necessary. For a large variety of materials such as GaN \cite{chen_homoepitaxial_2010,bergbauer_continuous-flux_2010,alloing_polarity_2011,brubaker_effect_2011,hestroffer_polarity_2011}, GaAs \cite{ikejiri_mechanism_2007}, or InAs \cite{tomioka_control_2008}, the crystal polarity has a crucial role in both the nucleation and the growth of nanowires. Concerning ZnO, the role of crystal polarity
was demonstrated already in 1972 by chemical etching for ZnO microwires
grown by chemical vapor deposition (CVD): the microwires were oriented
in the +\textbf{c} direction of the wurtzite structure (\textit{i.e.}
the Zn-polar $\left[0001\right]$ direction)\cite{iwanaga_note_1972,iwanaga_correction_1972}.
The Zn-polarity of ZnO nanoribbons was confirmed by convergent beam
electron diffraction (CBED) in 2003 \cite{wang_induced_2003}. A Zn-polarity
was also observed for nanowires grown by the hydrothermal method \cite{nicholls_polarity_2007},
by catalyst-assisted CVD \cite{jasinski_application_2008}, by pulsed
laser deposition (PLD) \cite{cherns_defect_2008-1}, and by evaporation
of zinc in an oxygen plasma \cite{baxter_growth_2003}, with the CBED
technique. Other methods such piezoelectric force microscopy have
shown the Zn-polarity of nanowires grown by the hydrothermal method
\cite{scrymgeour_polarity_2007}. Consequently, it was deduced that
the ZnO nanowire growth is due to a much higher growth rate along
the +\textbf{~c} direction. Results published about two-dimensional
thin layers agree with those about nanowires, with a higher growth
rate along the +\textbf{~c} direction than along the -\textbf{~c}
direction (the O-polar $\left[000\bar{1}\right]$ direction) \cite{wang_polarity_2005}.

Concerning the growth of ZnO nanowires on sapphire by PLD, Cherns
\emph{et al.} observed an O-polar under-layer but Zn-polar nanowires
\cite{cherns_defect_2008-1,sun_reduction_2008}. They showed that the nucleation of Zn-polar
nanowires occurred directly on the sapphire substrate, and suggested
that nuclei of both polarities could appear on sapphire which does
not present a polar character. However, this interesting mechanism
does not explain the nucleation of nanowires on ZnO substrates \cite{thiandoume_morphology_2009}.
As metalorganic vapor phase epitaxy (MOVPE) is an industrially-compatible
method which allows making hetero-structures with fast deposition
rates and a high reliability, a lot of work has been published concerning
the MOVPE growth of ZnO nanowires on sapphire \cite{jeong_comparative_2004,park_early_2005,cong_one-step_2005,park_surface_2006,park_defects_2007,behrends_investigation_2009,rosina_morphology_2009,perillat-merceroz_mocvd_2010}
since the first publication by G.C. Yi's group \cite{park_metalorganic_2002}.
In these publications, nanowires are obtained on a spontaneously deposited
ZnO under-layer, and sometimes the nanowires sit on top of ZnO pyramids.
The influence of stress is invoked in some publications to explain
the nanowire nucleation, because their larger surface-to-volume ratio
would help release the misfit strain between sapphire and ZnO \cite{cong_one-step_2005,park_surface_2006,liao_effect_2008,cao_tuning_2010}.
But this mechanism cannot account for the homo-epitaxial growth of
ZnO nanowires on ZnO substrates. In the case of MOVPE growth,
the role of the crystal polarity in the growth and nucleation mechanisms of ZnO nanowires, although mentioned in Ref.\cite{perillat-merceroz_mocvd_2010}, has not been discussed thoroughly. 

In this work, we aim to elucidate the role that polarity plays in
defining the morphology of MOVPE grown nanostructures. A comparison
is carried out on three different substrates: sapphire, O-polar ZnO,
and O-polar thick ZnO buffer layer grown previously on sapphire. For
the growth on sapphire, the normally reported morphology with an unintentional
under-layer, pyramids and nanowires is observed. For the direct growth
on O-polar ZnO (either on intentional buffer layers or on bulk substrates),
pyramids and nanowires are also observed. In every case, the crystal
polarity is determined by CBED, allowing us to link the shape of the
nanostructures with their polarity: whatever the substrate, nanowires
are found to be Zn-polar whereas the under-layers and the pyramids
are O-polar. Inversion domain boundaries (IDBs) are observed by transmission
electron microscopy (TEM), providing important indications on the
nucleation mechanisms. Observations of numerous nanowires for each
kind of sample allowed us to draw a clear picture of the nucleation
mechanisms involved for the different substrates used. On sapphire,
it is proposed that the non-polar character and the atomic steps of
this substrate are the cause for the germination of Zn-polar nanowires.
For the nucleation of Zn-polar ZnO on O-polar ZnO, the role of aluminum
impurities is discussed.

\section{Experimental details}

Three types of ZnO nanowire samples have been studied. Growth was
carried out either directly on $\left(0001\right)$ sapphire substrates,
or on a 400~nm thick buffer layer previously deposited on $\left(0001\right)$
sapphire substrates, or on O-polar hydrothermally-grown bulk ZnO substrates
(provided by Crystec). N\textsubscript{2} was used as the carrier
gas, N\textsubscript{2}O as the oxygen precursor and diethylzinc
as the zinc precursor. The pressure in the reactor was around 100~mbar.
The molar ratio between the O precursor and the Zn precursor allowed
controlling the morphology of the deposition: this ratio was about
500 for nanowire growth, and about 25000 for the two-dimensional buffer
layer growth. The growth temperature was between 750 and 850\textdegree{}C
for the nanowires, and of 950\textdegree{}C for the buffer layer.
A reactor without any rotating stage was used. Consequently, there
is a slight variation of the nanowire sizes from the center to the
edge of the substrate.

Cross-sectional TEM samples were prepared either by the cleaved edge
method \cite{den_hertog_mapping_2009}, or by mechanical polishing
followed by ion milling. The two-beam technique with $g=\left(0002\right)$
diffracting conditions was used to delineate domains with different
polarities. Convergent beam electron diffraction (CBED) patterns acquired
along a $\left\langle 1\bar{1}00\right\rangle $ zone axis were used
to determine the crystal polarity. To determine the orientation of
the CBED patterns relatively to the image, the sample is mechanically
lowered from the focal point in order to obtain a shadow image in
the diffraction disk without 180\textdegree{} rotation. Simulations
performed with the JEMS software \cite{stadelmann_jems_2004} were
used in order to index the $\left(0002\right)$ and the $\left(000\bar{2}\right)$
directions on the CBED experimental patterns, and consequently to
deduce the nanostructure polarities. Scanning TEM (STEM) images were
acquired with a high-angle annular dark field detector to localize
the position of the beam used for CBED. High-resolution TEM images
were acquired along $\left\langle 11\bar{2}0\right\rangle $ zone
axis.

\section{Results}

\subsection{Morphology of the nanostructures\label{sub:Morphology}}

\begin{figure}
\begin{centering}
\includegraphics[scale=0.65]{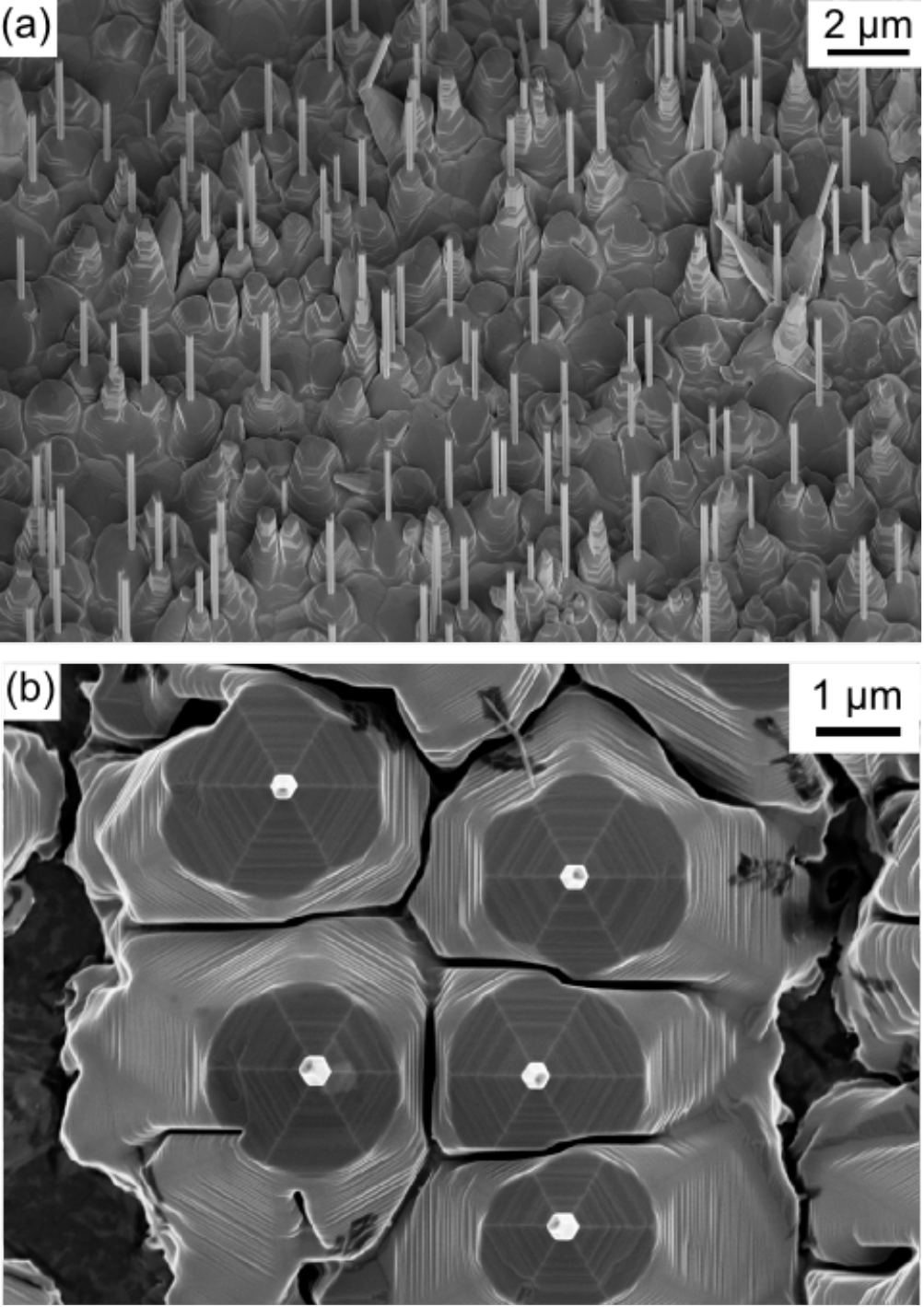}
\par\end{centering}

\caption{SEM images of ZnO nanowires and pyramids, grown during 45~min on
$\left(0001\right)$ sapphire: (a) tilted view, (b) top view.
}

\centering{}\label{Flo:NF tiltes MEB}
\end{figure}

The nanowire morphology for the growth on sapphire was described in
two previous papers \cite{rosina_morphology_2009,perillat-merceroz_mocvd_2010}.
We recall here the main results. Figure \ref{Flo:NF tiltes MEB} presents
scanning electron microscopy (SEM) images of vertical nanowires grown
on sapphire for a 45~min deposition time. They all present a pyramidal
base, even if some pyramids have no nanowire at their top. From x-ray
diffraction rocking-curves, the $\left(002\right)$ ZnO planes and
the $\left(006\right)$ sapphire planes were found to be parallel,
showing that the ZnO nanowires grew along the \textbf{c} direction
\cite{rosina_morphology_2009}. On the top-view SEM image (figure
\ref{Flo:NF tiltes MEB} (b)), it is seen that the nanowires have
a hexagonal section, and that they all exhibit the same in-plane orientation
with respect to sapphire. Two in-plane orientation relationships are
usually found for ZnO grown on $\left(0001\right)$ sapphire: in-plane
rotations of 0\textdegree{} or 30\textdegree{} can be observed \cite{liu_30_2006}.
Here, only the 30\textdegree{} rotation is observed, that is to say
the orientation relationship is $\left[0001\right]_{ZnO}$//$\left[0001\right]_{sapphire}$,
and $\left\langle 1\bar{1}00\right\rangle _{ZnO}$ //$\left\langle 11\bar{2}0\right\rangle _{sapphire}$
\cite{perillat-merceroz_mocvd_2010}. The nanowires exhibit $\left\{ \bar{1}100\right\} $
lateral faces, and do not contain any structural defects such as dislocations
or stacking faults \cite{perillat-merceroz_mocvd_2010}.

\begin{figure}
\begin{centering}
\includegraphics[scale=0.6]{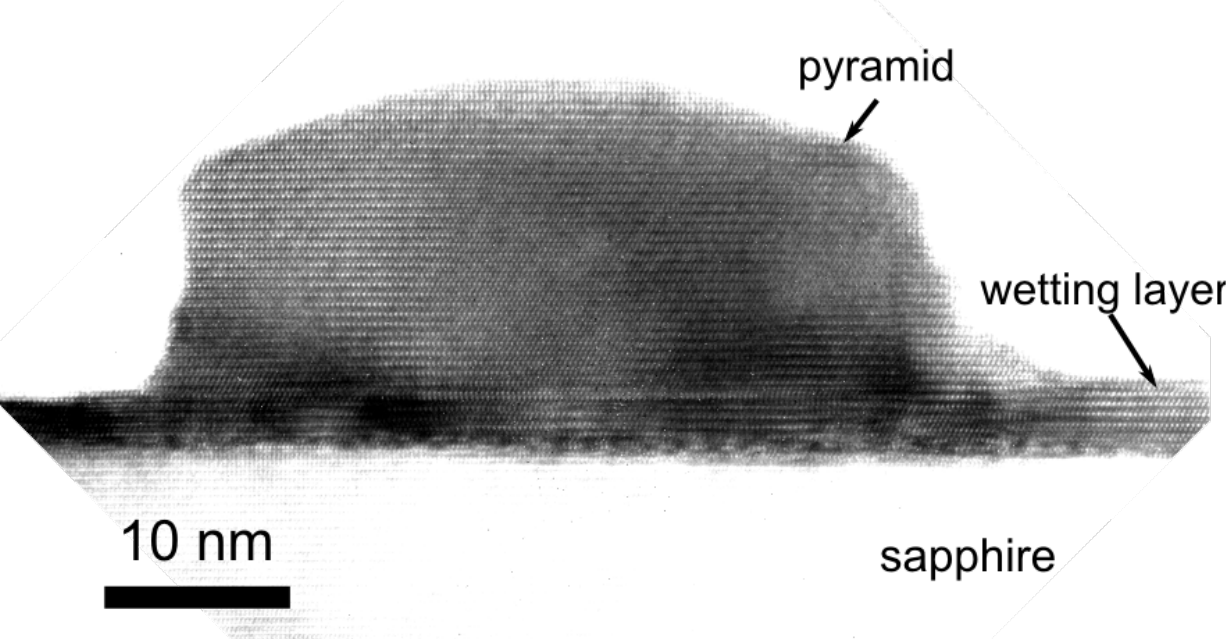}
\par\end{centering}

\caption{High resolution TEM image of a ZnO pyramid (without any nanowire) on the ZnO wetting layer, on the sapphire substrate, after one minute
growth.}

\centering{}\label{MET HR NF couche}
\end{figure}

As it was observed in all publications about growth of ZnO nanowires
on sapphire, a ZnO wetting layer is unintentionally formed \cite{park_metalorganic_2002,jeong_comparative_2004,park_early_2005,cong_one-step_2005,park_defects_2007,behrends_investigation_2009,rosina_morphology_2009,perillat-merceroz_mocvd_2010}.
It is present at the very beginning of the growth, even after a one
minute growth it is visible on the high resolution TEM image in figure
\ref{MET HR NF couche}.

\begin{figure}
\begin{centering}
\includegraphics[width=8cm]{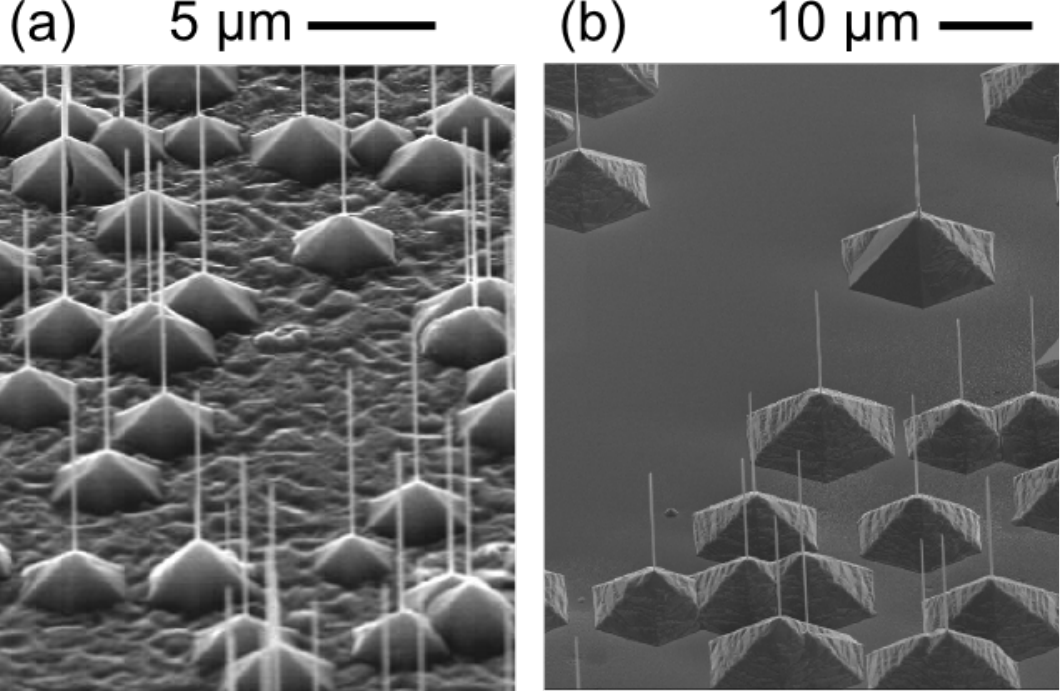}
\par\end{centering}

\caption{SEM images of ZnO nanowires and pyramids, grown during (a) 45~min
on a ZnO buffer layer on $\left(0001\right)$ sapphire, and (b) during
25~min on $\left(000\bar{1}\right)$ ZnO.}

\centering{}\label{Flo: MEB NF ZnO sur ZnO}
\end{figure}

Figure \ref{Flo: MEB NF ZnO sur ZnO} presents scanning electron microscopy
(SEM) images of vertical nanowires grown on a ZnO buffer layer deposited
on $\left(0001\right)$ sapphire (Figure \ref{Flo: MEB NF ZnO sur ZnO}
(a)), and on a $\left(000\bar{1}\right)$ ZnO substrate (Figure \ref{Flo: MEB NF ZnO sur ZnO}
(b)). As for the growth on sapphire, all nanowires present a pyramidal
base.

\subsection{Polarity determination\label{sub:Polarity-determination}}

\begin{figure*}
\begin{centering}
\includegraphics[scale=0.53]{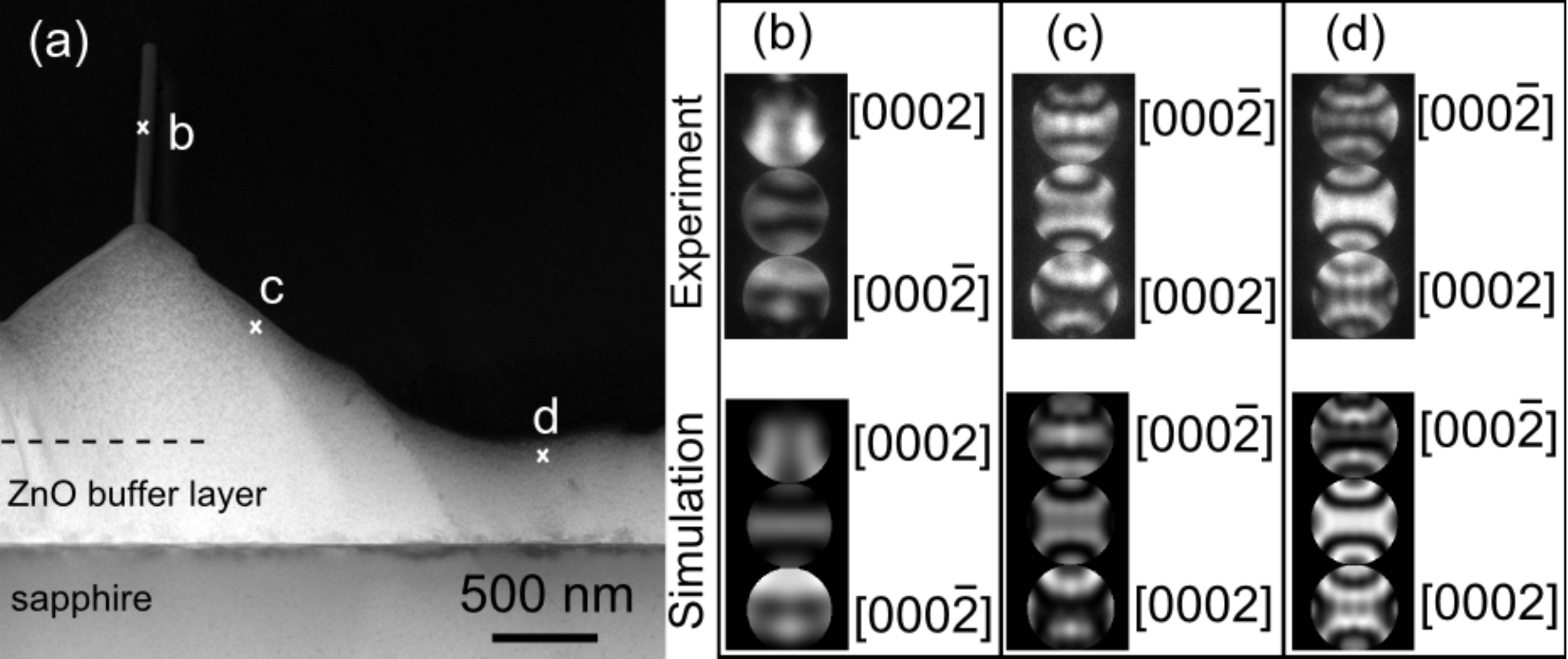}
\par\end{centering}

\caption{(a) STEM image of a ZnO nanowire on top of a ZnO pyramid on a ZnO
buffer layer, on the sapphire substrate. (b)(c)(d) Experimental (top)
and simulated (bottom) CBED patterns of the nanowire, the pyramid,
and the buffer layer, respectively. Simulation for ZnO thicknesses
of 46, 110, and 140~nm respectively, done with the JEMS software
\cite{stadelmann_jems_2004}. For all CBED patterns, the growth direction
is upwards. The nanowire is Zn-polar, the pyramid and the buffer layer
are O-polar.}

\centering{}\label{Flo:CBED ZnO buffer}
\end{figure*}

The polarity of ZnO nanowires and pyramids was determined for two
types of samples, one grown on a ZnO buffer layer, and the other grown
directly on sapphire. Figure \ref{Flo:CBED ZnO buffer} (a) is a STEM
image of a nanowire on top of a pyramid, on the 400~nm thick ZnO
buffer layer previously deposited on sapphire.%
{} The CBED patterns corresponding to the nanowire are shown in figure
\ref{Flo:CBED ZnO buffer} (b) with the experimental pattern on top,
and the simulated one at the bottom. The growth direction is along
the $\left[0002\right]$ direction: in other words, the nanowire is
Zn-polar. The CBED experimental and simulated patterns (top and bottom
respectively) corresponding to the pyramid are shown in figure \ref{Flo:CBED ZnO buffer}
(c). The growth direction is along the $\left[000\bar{2}\right]$
direction: pyramids are O-polar. Similarly, the CBED patterns of figure
\ref{Flo:CBED ZnO buffer} (d) reveals the O polarity of the ZnO buffer
layer. Thus, we might infer that opposite crystal polarities lead
to very different growth rates: Zn-polar crystals lead to nanowires
while O-polar crystals lead to pyramids, for the conditions used here
for nanowire growth. Ab-initio\emph{ }\textit{\emph{calculations help
to explain this strong anisotropy, which is observed whatever the
elaboration method is. The cohesion energies of Zn and O atoms on
the two $\left\{ 0001\right\} $ surfaces were calculated \cite{na_first-principles_2010}.
For a Zn-O dimer, they are similar for the two surfaces. However,
the zinc cohesion energy is much more important on the $\left(0001\right)$
Zn-polar surface (-6.9~eV) compared to the $\left(000\bar{1}\right)$
O-polar surface (-1.3~eV): this could explain the strong growth anisotropy
along the $\pm$}}\textbf{\textit{\emph{~c}}}\textit{\emph{ directions.}}\emph{
}

\begin{figure}
\begin{centering}
\includegraphics[width=8cm]{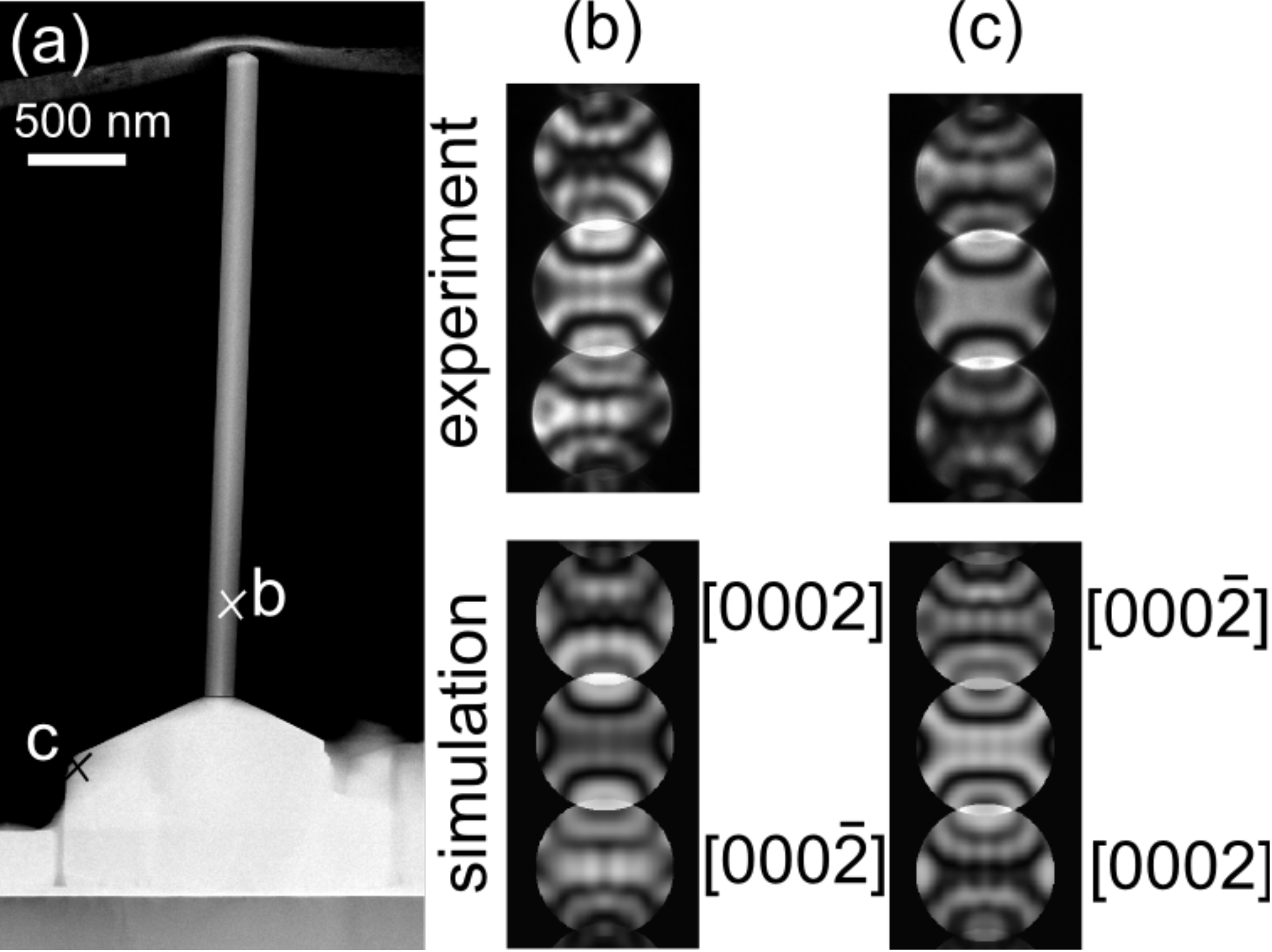}
\par\end{centering}

\caption{
(a) STEM image of a ZnO nanowire and a ZnO pyramid on a unintentional
ZnO wetting layer on sapphire. Experimental (up) and simulated (down)
CBED patterns (b) of the nanowire, and (c) of the pyramid, with the
growth direction upwards. Simulations with the JEMS software for thicknesses
of 100 and 113~nm \cite{stadelmann_jems_2004}.
}

\centering{}\label{Flo:CBED ZnO saphir}
\end{figure}

CBED experiments were also done for nanowires and pyramids grown directly
on sapphire, with the presence of an unintentional ZnO wetting layer
(figure \ref{Flo:CBED ZnO saphir}). Once again, nanowires are Zn-polar
and the pyramids are O-polar. It was not possible to determine the
polarity of the thin wetting layer, maybe because of the presence
of numerous structural defects. However, the same morphology with
pyramids and nanowires is obtained on this thin wetting layer, and
on the O-polar buffer layer and on O-polar hydrothermal substrates.
This is in contrast with the growth on Zn-polar substrates, for which
only a thin two-dimensional layer was obtained instead of nanowires
for the same growth conditions (result not shown here). Consequently
we assume that this wetting layer is also O-polar. This dependence
of the nanostructure shape on the crystal polarity seems to be general
for wurtzite materials. For example, it was shown for MOVPE-grown
GaN nanostructures that pyramids were Ga-polar and nanowires N-polar
\cite{chen_homoepitaxial_2010,bergbauer_continuous-flux_2010,alloing_polarity_2011}.

\subsection{Inversion domain boundaries : position and formation }

After having determined that pyramids were O-polar and nanowires were
Zn-polar, the position and shape of the interface between these inverted
domains has been studied by two-beam TEM. The results for the growth
on O-polar substrates or on O-polar buffer layer are reported in this
section, and those concerning the growth directly on sapphire in the
following one.

\begin{figure}
\begin{centering}
\includegraphics[scale=0.6]{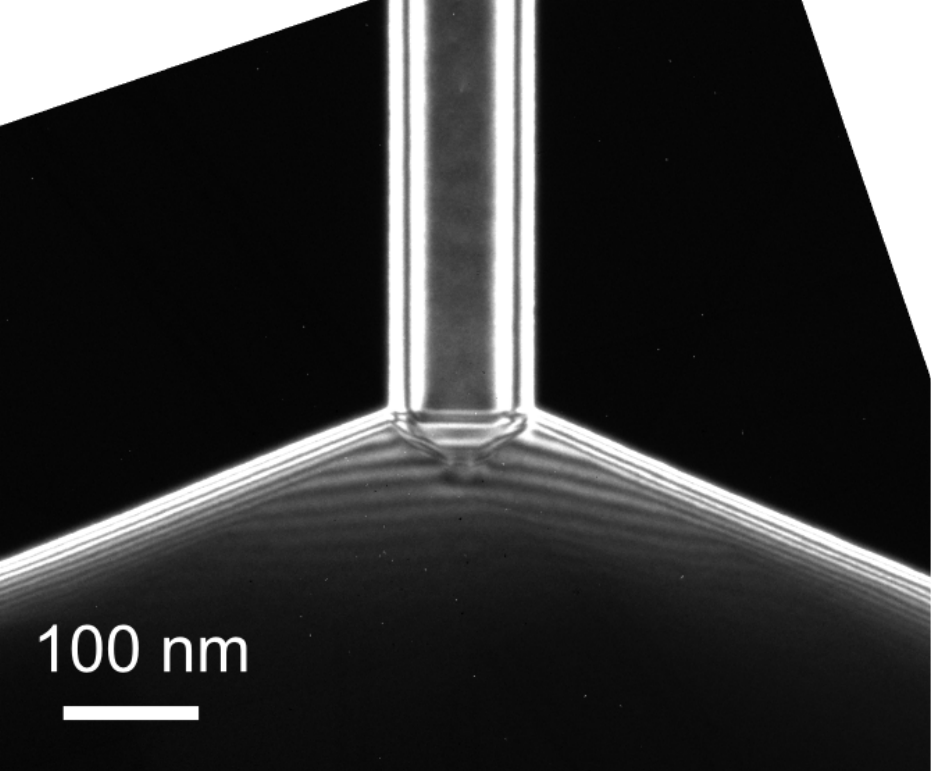}
\par\end{centering}

\caption{
Two-beam TEM image with $g=\left(0002\right)$ of a Zn-polar nanowire
on a O-polar pyramid, grown during 45~min on a ZnO buffer layer on
sapphire. The IDB between the nanowire and the pyramid is at the top
of the pyramid, as observed in 90\% of cases for the growth on O-polar
ZnO.
}

\centering{}\label{Flo:Nf sommet pyr}
\end{figure}

\begin{figure}
\begin{centering}
\includegraphics{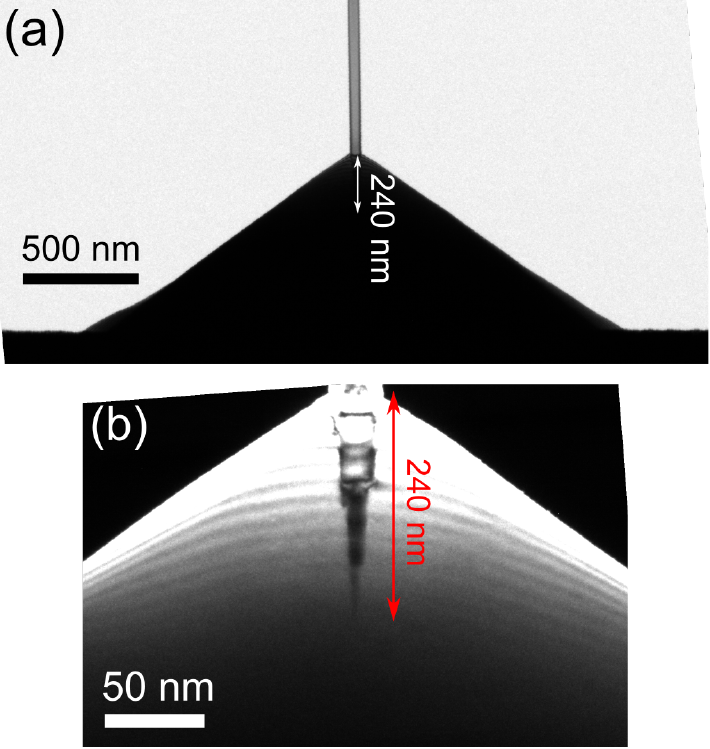}
\par\end{centering}

\caption{
Two-beam TEM images with $g=\left(0002\right)$ of a Zn-polar nanowire
and a O-polar pyramid grown on an O-polar ZnO substrate. The Zn-polar
domain of the nanowire continues under the top of the pyramid, which
is observed in 10\% of cases for the growth on O-polar ZnO.
}

\centering{}\label{Flo:NF milieu pyr}
\end{figure}

\begin{table*}
\begin{centering}
\begin{tabular}{|c|c|c|c|c|}
\hline 
Substrate & Deposition & IDB at the top  & IDB inside  & IDB at the interface \tabularnewline
 & time & of the pyramid & the pyramid  & with the substrate\tabularnewline
\hline 
Sapphire + buffer ZnO & \textcolor{black}{45~min} & 10/10 & 0/10 & 0/10\tabularnewline
ZnO & 5~min & 7/7 & 0/7 & 0/7\tabularnewline
ZnO & 5~min & 17/22 & 5/22 & 0/22\tabularnewline
ZnO & 25~min & 11/11 & 0/11 & 0/11\tabularnewline
Sapphire & 30~min & 8/18 & 2/18 & 8/18\tabularnewline
\hline 
\end{tabular}
\par\end{centering}

\caption{
Summary table of the position of the IDBs according to the samples.
}

\centering{}\label{Flo:Tableau position ID}
\end{table*}

Figure \ref{Flo:Nf sommet pyr} shows a $g=\left(0002\right)$ TEM
image of a nanowire on a pyramid, grown on an O-polar ZnO buffer layer
on sapphire. Clearly, the Zn-polar domain begins at the top of the
O-polar pyramid. The same result was obtained for all of the ten nanowires
and pyramids observed on this sample. These observations were repeated
for nanowires and pyramids grown during five minutes on a O-polar
ZnO substrate. For all of the seven observations, the IDBs are also
found to be lying at the top of the pyramid. Two other samples were
grown on ZnO substrates with exactly the same growth conditions, but
with different deposition times of 5 and 25~min. On the 25 min sample,
the IDBs were found at the top of the pyramids for all of the 11 observations.
This is in contrast with the sample grown for 5 min, where, for 5
nanowires over 22, the Zn-polar domain seems to originate from a certain
depth within the pyramids but never from the interface with the substrate
(figure \ref{Flo:NF milieu pyr}). For the 17 other wires, the IDBs
lie at the top of the pyramid as usually observed. The statistics
of these observations are summarized in table \ref{Flo:Tableau position ID}.
Considering the four samples grown on O-polar ZnO (one growth on a
buffer layer, and the three other growths on bulk substrates), it
is deduced that in 90\% of cases, the Zn-polar domain sits at the
top of the O-polar pyramid. It is interesting to notice that the IDBs
present inclined facets rather than basal ones: this point will be
discussed later.

\begin{figure}
\begin{centering}
\includegraphics[scale=0.5]{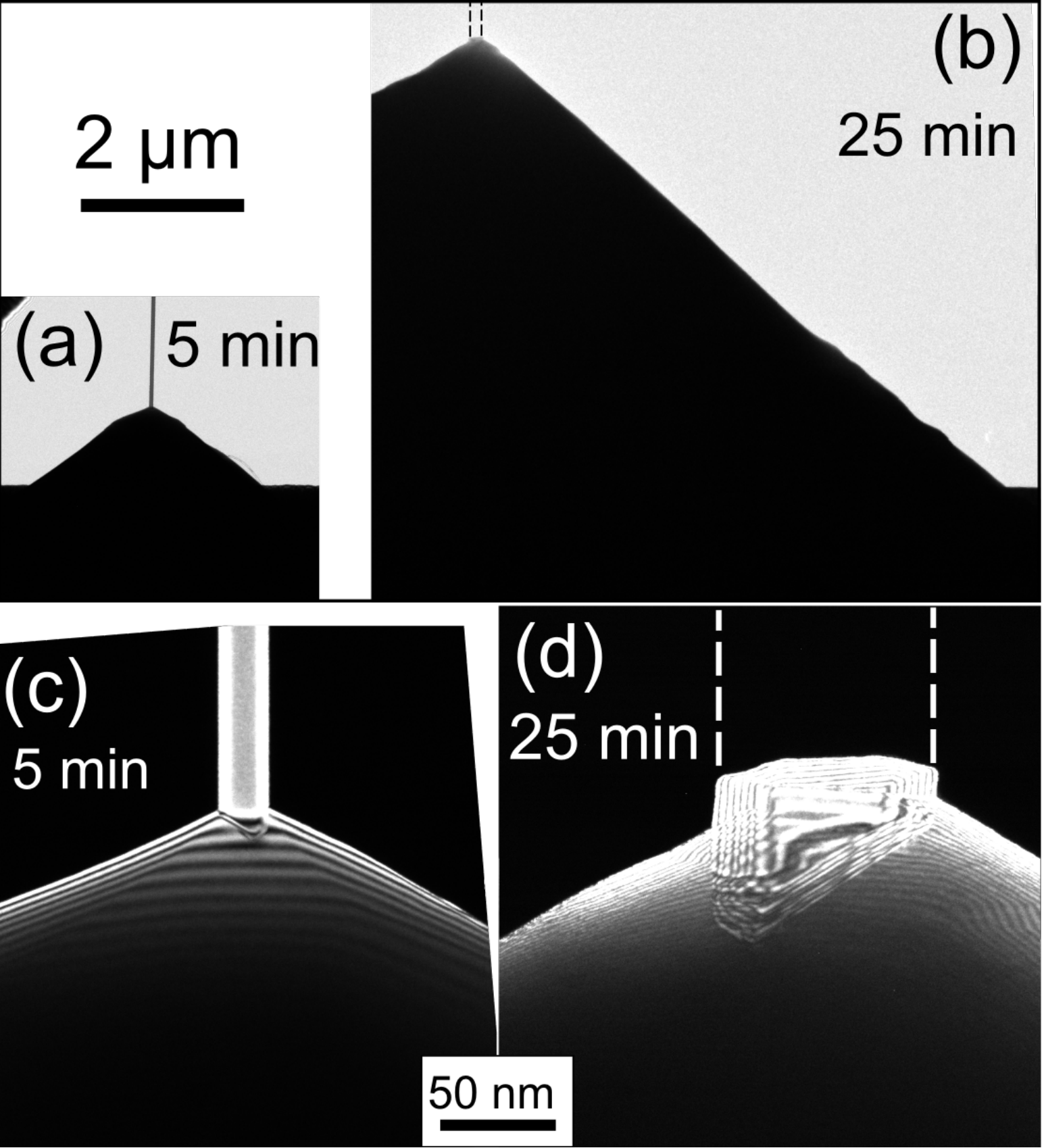}
\par\end{centering}

\caption{
Two-beam TEM images with $g=\left(0002\right)$ of Zn-polar nanowires
on top of O-polar pyramids, grown on an O-polar ZnO substrate, for
(a) (c) 5~min of growth and (b) (d) 25~min of growth. (a) (b): low
magnification images at the same scale. (c) (d): zooms on the IDBs
at the tops of the pyramids. The dashed lines on images (b) and (d)
recall the position of the nanowire removed during the TEM sample
preparation.
}

\centering{}\label{Flo:NF 300 1500s}
\end{figure}

Figure \ref{Flo:NF 300 1500s} images a nanowire on top of a pyramid
after a 5~min growth ((a) and (c)), and after a 25~min growth ((b)
and (d)). Figure \ref{Flo:NF 300 1500s} (a) and (b) are low magnification
images of ZnO pyramids at the same scale: pyramids have grown in height
and diameter with time. Similarly, nanowires have grown, even if the
variation in height is not visible on these images. By observations
of many pyramids and nanowires (22 observations for 5~min growth
and 11 observations for 25~min growth), we determined that the pyramid
heights after 5~min growth are between 0.5 and 1.3~\textmu{}m, with
a mean height of 0.8~\textmu{}m. Nanowire diameters are between 10
and 100~nm, with an average value of 50~nm. For a 25~min growth,
the pyramid heights are between 1 and 9~\textmu{}m, with an average
height of 5~\textmu{}m. Nanowire diameters are between 20 and 300~nm,
with an average diameter of 180~nm. Consequently, the pyramids and
nanowires presented in figure \ref{Flo:NF 300 1500s} (a) and (b)
are thought to be representative. The IDB is at the top of the pyramid
for the 5 min growth (figure \ref{Flo:NF 300 1500s} (c)), and it
stays at the top of the pyramid after 25 min of growth (figure \ref{Flo:NF 300 1500s}
(d)). It is deduced that the IDB moves up during the growth of the
pyramid and of the nanowire.

\begin{figure}
\begin{centering}
\includegraphics[scale=0.7]{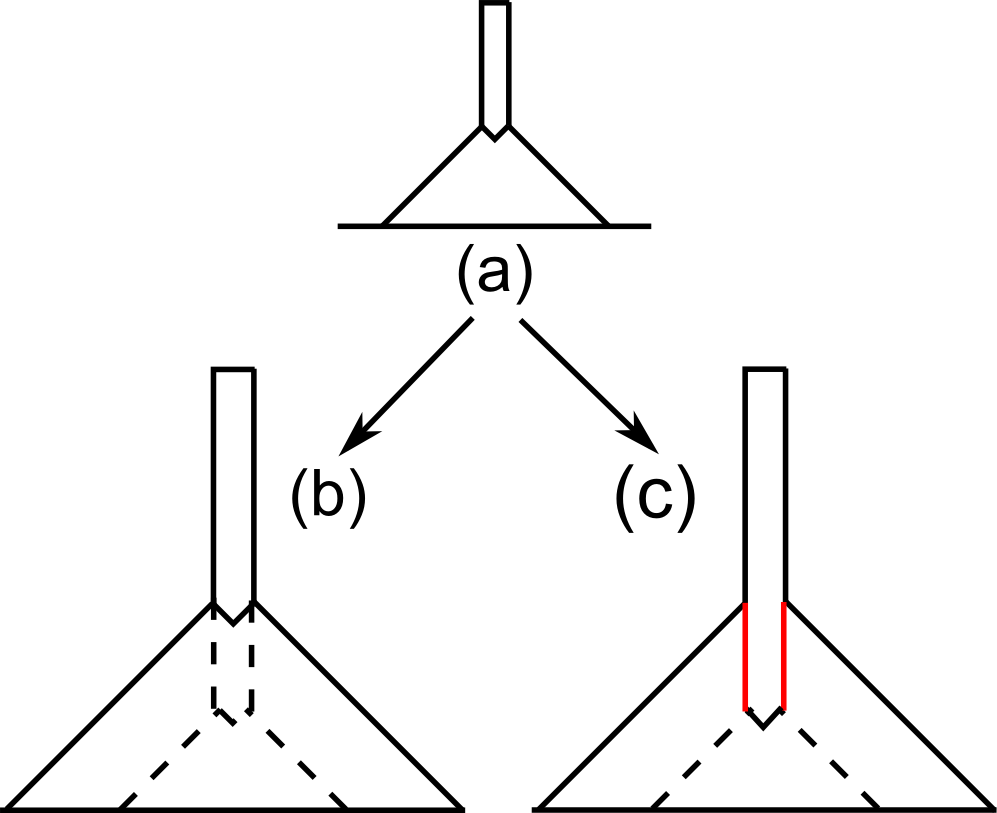}
\par\end{centering}

\caption{
(a) Scheme of a Zn-polar nanowire growth on top of a O-polar pyramid.
(b) The Zn-polar domain moves up during the growth and stays at the
top of the pyramid. (c) The Zn-polar domain does not move during growth.
}

\centering{}\label{Flo:Schema mouvement IDB}
\end{figure}

Figure \ref{Flo:Schema mouvement IDB} is a schematic of this IDB
movement. On figure \ref{Flo:Schema mouvement IDB} (a), the Zn-polar
nanowire is drawn on top of the O-polar pyramid. Figures \ref{Flo:Schema mouvement IDB}
(b) and (c) present the nanowire and the pyramid later during the
growth, in the case where the IDB moves up (figure \ref{Flo:Schema mouvement IDB}
(b)), and in the case where the IDB does not move. In this latter
case vertical IDBs (in red on figure \ref{Flo:Schema mouvement IDB}
(c)) are formed. We assume that in the general case (90\% of cases),
the energy necessary to move the IDB is smaller than the formation
energy of the vertical IDB on the side walls of the nanowires. Consequently,
the upward movement of the IDB would be favored. There is no data
in the literature about the ZnO IDB energies. One possibility would
be to determine the atomic structure of the inclined and vertical
IDBs from high-resolution TEM images, and then calculate the IDB energies
by \emph{ab-initio} calculations. But the sample preparation to obtain
a nanowire on a pyramid with a thickness less than 50 nm is very tricky.
We will discuss in subsection \ref{sec:Discussion:-nanowire-nucleation}
on the possible influence that segregating aluminum impurities coming
from the sapphire substrate or from the ZnO substrate could bear on
the formation of the IDBs. The displacement rate of the IDBs might
be correlated with the diffusion rate of Al, because impurities may
have a drag effect during grain boundary movement \cite{consonni_effects_2011}.
Different impurity concentrations may explain the fact that in 10\%
of the cases (for the growth on O-polar ZnO), the IDB is pinned inside
the pyramid. Moreover, the grain boundary movement is thermally activated
because it is linked to the diffusion of Zn and O atoms and of extrinsic
impurities. Thus studying the influence of the temperature and of
growth durations on the movement of the IDBs could yield valuable
informations on the activation energy for this movement. However,
the variation range of temperature is limited, because the nanostructure
morphology is temperature-dependent.

\begin{figure}
\begin{centering}
\includegraphics[scale=0.54]{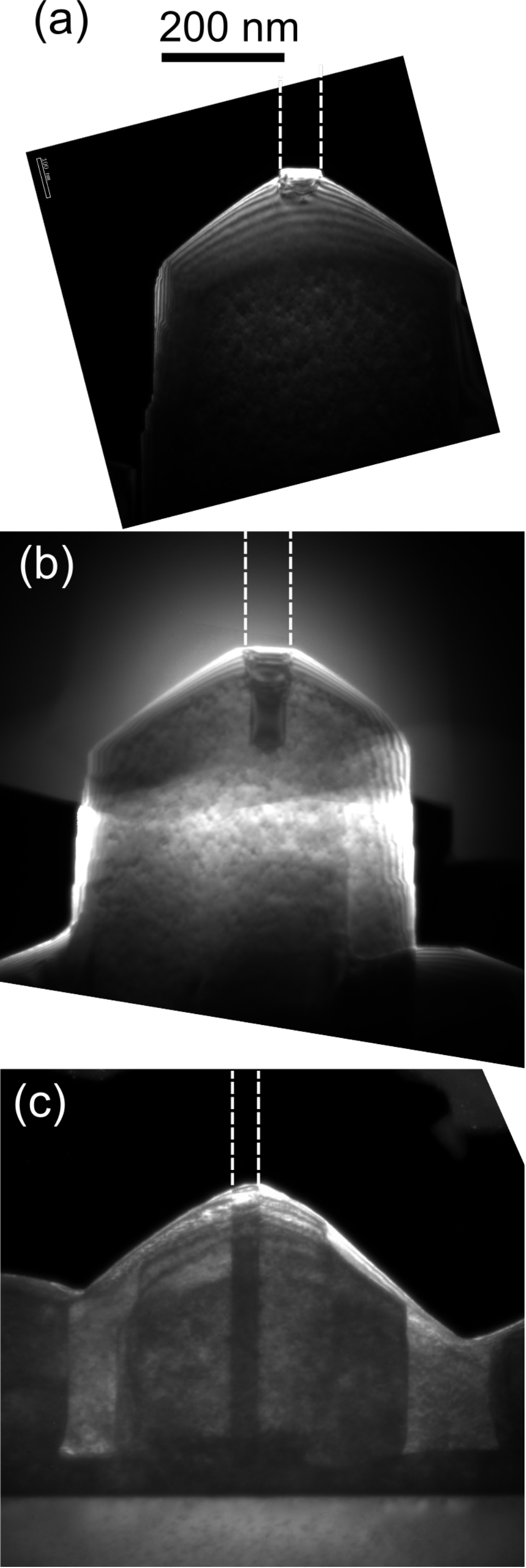}
\par\end{centering}

\caption{Two-beam TEM images with $g=\left(0002\right)$ of ZnO pyramids and
nanowires grown during 30 min directly on sapphire : (a) IDB at the
top of the pyramid, (b) IDB inside the pyramid, (c) IDB originating
from the sapphire / ZnO interface. The dashed lines recall the position
of the nanowires broken during the TEM sample preparation.
}

\centering{}\label{Flo:germination sur saphir 6349}
\end{figure}

For comparison, IDBs are examined for growth directly on sapphire.
The presence of an unintentional ZnO wetting layer, together with
pyramids and nanowires has been noticed earlier. Figure \ref{Flo:germination sur saphir 6349}
shows TEM images with $g=\left(0002\right)$ for three different nanowires
and pyramids grown for 30~min on sapphire, presenting three different
cases. For 8 observations over 18, the IDBs lie at the top of the
pyramids, as observed in 90\% of the cases for the growth on O-polar
ZnO (figure \ref{Flo:germination sur saphir 6349} (a)). For 2 of
the 18 observations, the IDBs originate from inside the pyramids,
as observed in 10\% of the cases for the growth on ZnO (figure \ref{Flo:germination sur saphir 6349}
(b)). Finally, for 8 of the 18 observations, the Zn-polar domains
originate at the sapphire / ZnO interface (figure \ref{Flo:germination sur saphir 6349}
(c)): this case was never observed for the growths on O-polar ZnO
(either buffer layer or bulk substrate). These statistics are also
summarized in table \ref{Flo:Tableau position ID}.

\section{Discussion: nanowire nucleation\label{sec:Discussion:-nanowire-nucleation}}

\subsection{Nanowire nucleation on O-polar ZnO}

The spontaneous nucleation of Zn-polar nanowires on O-polar ZnO has
never been reported in the literature. It is interesting to compare
with the GaN case, which has the same wurtzite structure as ZnO. In
this discussion, the literature concerning GaN is briefly reviewed,
then the role of impurities related to the polarity inversions in
ZnO is examined.

\begin{figure}
\begin{centering}
\includegraphics{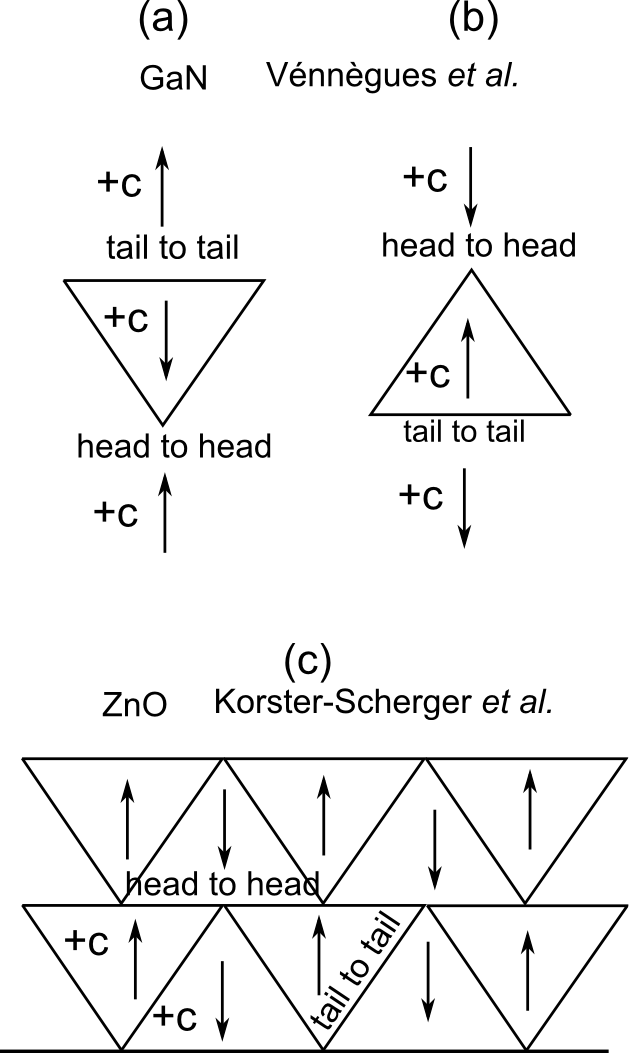}
\par\end{centering}

\caption{
Configuration of the inverted domains for (a), (b) Mg-doped GaN \cite{vennegues_pyramidal_2000},
and (c) Fe\textsubscript{2}O\textsubscript{3}-doped ZnO \cite{korster-scherger_zno_2007}.
Arrows indicate the +\textbf{c} direction. (a) N-polar domain in a
Ga-polar matrix (b) Ga-polar domain in a N-polar matrix. (a) and (b):
for GaN, configuration head-to-head for the inclined IDBs, and tail-to-tail
for the basal IDBs. (c) For ZnO, configuration tail-to-tail for the
inclined IDBs, and head-to-head for the basal IDBs.
}

\centering{}\label{Flo:tete a tete}
\end{figure}

For GaN, spontaneous polarity inversions during growth have never
been observed. However, doping with Mg is often correlated with the
formation of polarity inversion domains during the growth of thin
two-dimensional layers \cite{ramachandran_inversion_1999,vennegues_pyramidal_2000,vennegues_atomic_2003}.
For MOVPE growth, small triangular domains with an inverted polarity
compared to the matrix are observed because of Mg-doping. In these
samples, the average Mg concentration is estimated to be 10\textsuperscript{19}cm\textsuperscript{-3}.
The shape of the triangular domains is shown in figure \ref{Flo:tete a tete}
(a) for a matrix growing in the +\textbf{~c} direction (Ga polarity),
and in figure \ref{Flo:tete a tete} (b) for a matrix growing in the
-\textbf{~c} direction (N polarity). Inclined IDBs are observed for
the so-called head-to-head configuration, whereas basal IDBs are observed
for the tail-to-tail configuration \cite{vennegues_pyramidal_2000}.
Concerning GaN nanostructures grown by MOVPE, nanowires are N-polar
and pyramids Ga-polar. However, nucleation of N-polar nanowires occurred
either directly on sapphire, or on N-polar GaN covered by a thin SiN\textsubscript{x}
mask, but never on Ga-polar pyramids \cite{chen_homoepitaxial_2010}.

Concerning two-dimensionnal ZnO layers, it was observed that their
polarity depended on Al doping. For ZnO growths without doping on
sapphire \cite{adachi_change_2008} or on glass \cite{adachi_polarity_2011},
O-polar ZnO was obtained. However, when ZnO was doped with 1\% Al,
a Zn-polar layer is obtained. For application in varistors, polycrystalline
ZnO is synthesized by sintering of ZnO powders with the addition of
others oxides such as SnO\textsubscript{2} \cite{recnik_nucleation_2007}
or Fe\textsubscript{2}O\textsubscript{3} \cite{korster-scherger_zno_2007}.
These dopants are known to form spinel phases with ZnO and to trigger
the formation of IDBs in the ZnO crystals. It is deduced that for
doped ZnO, a microstructure with IDBs is thermodynamically more stable
than a solid solution of dopants in ZnO. Models are proposed for the
formation of basal plane IDBs \cite{recnik_nucleation_2007}, and
inclined plane IDBs \cite{korster-scherger_zno_2007}. According to
these works, dopants can diffuse into ZnO, and segregate to octahedral
sites of the ZnO hexagonal structure. To preserve the local charge
balance, Zn atoms are displaced to form inverted domains. IDBs with
basal planes are created in the case of impurities with an oxidation
state greater than three such as Sn, while IDBs with basal and inclined
planes are created in the case of impurities with an oxidation state
equal to three such as Fe.

For Fe-doped ZnO, the shape of the IDBs is shown in figure \ref{Flo:tete a tete}
(c): IDBs lie in the basal planes for a head-to-head configuration,
and in inclined planes for a tail-to-tail configuration, contrary
to the Mg-doped GaN case. In our case, where Zn-polar nanowires sit
on top of O-polar pyramids, the configuration is tail-to-tail: our
observations of inclined IDBs are thus consistent with the observations
of Korster-Scherger \emph{et al.} (figure \ref{Flo:tete a tete} (c)).
Either for GaN or for ZnO, it was reported in the literature that
impurities segregate to the IDBs when relatively large quantities
of dopants are intentionally introduced \cite{vennegues_atomic_2003,mccoy_inversion_1996,daneu_grain_2003}.
In our case, Al impurities might be the cause for the nucleation
of Zn-polar domains. These impurities could diffuse from the sapphire
or from the hydrothermally grown ZnO substrates, the concentration
of Al in the ZnO substrate measured by secondary ion mass spectroscopy
being around 5.10\textsuperscript{17} cm\textsuperscript{-3}. But
Al was not detected on the IDBs by electron dispersive x-ray spectroscopy
(EDX), probably because the concentration is too low.

In order to prove the role of Al in the nucleation of polarity inversion,
it would be interesting to grow nanowires and pyramids simultaneously
on purer ZnO substrates (for example, those provided by Tokyo Denpa
with an Al concentration of about 5.10\textsuperscript{15} cm\textsuperscript{-3}),
on the ZnO substrate used here (provided by Crystec with an Al concentration
of about 5.10\textsuperscript{17} cm\textsuperscript{-3}), and on
an intentionally doped ZnO layer (with an Al concentration of about
10\textsuperscript{19} cm\textsuperscript{-3}). For a higher aluminum
doping, we would expect a higher nanowire density, and Al might be
detected by EDX (or another method) on the IDBs.

\subsection{ZnO nanowire nucleation on sapphire}

Contrary to the nucleation of ZnO nanowires on ZnO substrates, direct
nucleation on sapphire is well documented. Three hypothesis are commonly
proposed in the literature, and discussed below. A majority of publications
describing the MOVPE growth of ZnO nanowires on sapphire invokes the
role of stress to explain the nucleation \cite{cong_one-step_2005,park_surface_2006,liao_effect_2008,cao_tuning_2010}.
Nanowires would help to release the stress in the ZnO under-layer
thanks to their free surfaces. This mechanism may be valid, but it
is uncomplete because it does not take into account the fact that
the under-layer and the nanowires have different polarities. Furthermore,
as discussed previously, nanowires are found to nucleate as well on
ZnO substrates: the stress hypothesis can be ruled out in this case.
Finally, it does not explain the nucleation at the sapphire ZnO interface.

The role of atomic steps in the nucleation of IDBs was proposed first
for GaN \cite{ruterana_growth_2000}. The nucleation of an IDB above
an atomic step helps accommodating the misfit due to the vertical
translation of the two crystals on each side of the atomic step. For
ZnO, IDBs nucleating at atomic steps were also observed \cite{park_defects_2007}.
In our case, we rarely observed an alignment of the nanowires along
the sapphire crystallographic directions \cite{perillat-merceroz_mocvd_2010}.
Consequently, we think that atomic steps are only one possibility
to initiate polarity inversions with the subsequent formation of nanowires.

Another possibility for the nanowire nucleation is proposed by Cherns
\emph{et al} \cite{cherns_defect_2008-1}\emph{.} Based on their TEM
observations, they proposed that the Zn-polar domains which give rise
to nanowires during the PLD growth on sapphire directly originated
from the sapphire surface. This would be due to the initial growth
conditions. They recall that ZnO of both polarities can be obtained
in a controlled manner depending on the sapphire surface treatments
\cite{mei_controlled_2004,mei_controlled_2005}. Moreover, polarity
inversions originating at the ZnO / sapphire interface were observed
during the growth of thin layers \cite{narayan_defects_1998} on sapphire.
Actually, because sapphire is non polar, it is not surprising that
both polarities can be obtained. In our case, the wetting layer of
O-polarity would result from the lateral growth of the O-polar nuclei,
with a few Zn-polar nuclei inclusions. These Zn-polar nuclei would
give rise to Zn-polar nanowires because of the strong anisotropy between
the growth rates of O-polar and Zn-polar ZnO.

\section{Summary}

The nucleation mechanisms of ZnO nanowires were investigated for different
types of substrates, namely c-oriented sapphire and O-polar ZnO. The
MOVPE-grown nanostructures mainly consist of pyramids with nanowires
sitting on top. It was shown by CBED that, whatever the substrate,
nanowires and pyramids have opposite polarities: nanowires are Zn-polar
and pyramids are O-polar. The examination of the inversion domain
boundaries (IDBs) for a fair population of nanowires grown either
in homo-epitaxy on O-polar ZnO templates or substrates, or in hetero-epitaxy
on sapphire, allowed us to draw a clear picture of the nucleation
mechanisms at play in the two cases. For growth on O-polar ZnO, nanowire
nucleation implies the creation of an inverted domain. Its origin
is discussed and possibly attributed to the presence of aluminum impurities
diffusing from the substrates. This hypothesis would need to be comforted
by comparing nanowires grown on ZnO substrates with different residual
aluminum concentrations and on aluminum doped ZnO templates. The IDBs
formed between the O-polar pyramid and the Zn-polar nanowire was shown
to move during the growth, and to be always located at the top of
the pyramid. For growth on sapphire substrates, the nucleation of
Zn-polar domains occurred either on the O-polar pyramids, or directly
on sapphire. In this last case, nucleation is attributed partly to
atomic steps, but mainly to the non-polar character of sapphire. To
conclude, this work stresses the fact that crystal polarity governs
both the nucleation and the shape of ZnO nanostructures. Moreover,
the presence of IDBs is shown in
self-assembled nanowires grown on sapphire and ZnO. A particular attention
should be paid to these defects which could be detrimental to the
device efficiencies.

\ack
The authors acknowledge funding from the French national research agency (ANR) through the Carnot program (2006/2010).

\section*{References}
\bibliographystyle{unsrt}
\bibliography{bibliotheque}

\end{document}